\documentclass[prd,aps,floats,preprint,preprintnumbers]{revtex4}

\usepackage{amsmath}
\usepackage{amssymb}
\usepackage{graphicx}
\usepackage{psfrag}

\def\bea{\begin{eqnarray}}
\def\eea{\end{eqnarray}}

\def\He{{\cal H}}
\def\a{\alpha}
\def\b{\beta}
\def\g{\gamma}
\def\d{\delta}

\def\g{\gamma}

\def\r{\rho}
\def\s{\sigma}
\def\t{\tau}
\def\lag{{\cal L}}

\def\N{\nabla}
\def\s{\sigma}

\def\t{\tau}
\newcommand{\half}{\frac12}
\def\pa{\partial}
\def\quarter{\frac{1}{4}}

\def\na{\nabla}
\def\al{\alpha} 
\def\be{\beta} 
\def\ga{\gamma}
\def\de{\delta}

\def\rh{\rho}
\def\si{\sigma}

\def\Ga{\Gamma}

\newcommand{\GB}{R^2_{\rm GB}}

\begin{document}

\setlength{\unitlength}{1mm}

\title{Ghosts, Instabilities, and Superluminal Propagation in Modified Gravity Models}

\author{Antonio De Felice$^1$\footnote{a.de-felice@sussex.ac.uk}, Mark Hindmarsh$^1$\footnote{m.b.hindmarsh@sussex.ac.uk}, Mark Trodden$^2$\footnote{trodden@physics.syr.edu}}

\affiliation{$^1$Department of Physics \&\ Astronomy, University of Sussex,
Brighton BN1 9QH, United Kingdom.\\
$^2$Department of Physics, Syracuse University,
Syracuse, NY 13244-1130, USA.}

\begin{abstract}
We consider Modified Gravity models involving inverse powers of fourth-order curvature invariants. Using these models' equivalence to the theory of a scalar field coupled to a linear combination of the invariants, we investigate the properties of the propagating modes. Even in the case for which the fourth derivative terms in the field equations vanish, we find that the second derivative terms can give rise to ghosts, instabilities, and superluminal propagation speeds.  We establish the conditions which the theories must satisfy in order to avoid these problems in Friedmann backgrounds, and show that the late-time attractor solutions generically exhibit superluminally propagating tensor or scalar modes.
\end{abstract}

\maketitle

\section{Introduction}
\label{sec:intro}
The strong observational evidence for an accelerating universe~\cite{Riess:1998cb,Perlmutter:1998np,Riess:2004nr,Tonry:2003zg,Spergel:2006hy,Page:2006hz,Hinshaw:2006ia,Jarosik:2006ib,Netterfield:2001yq,Halverson:2001yy} has sparked a widespread search for a dynamical explanation. Beyond a bare cosmological constant, a plethora of other models have been proposed, with quintessence - a dynamical scalar field that behaves essentially as a modern day inflaton field, - being perhaps the simplest example (see~\cite{Weiss:1987xa,Wetterich:1987fm,Ratra:1987rm,Peebles:1987ek}). In this context, many potentials have been introduced that yield late-time acceleration and tracking behaviour (see~\cite{Frieman:1995pm,Coble:1996te,Peebles:1998qn,Steinhardt:1999nw,Zlatev:1998tr,Ferreira:1997au,Liddle:1998xm,Copeland:1997et,Ng:2001hs,Bludman:2004az}).

Among other approaches, Modified Gravity Models have attracted great interest (see~\cite{Carroll:2003wy,Carroll:2004de,Capozziello:2003tk,Deffayet:2001pu,Freese:2002sq,Arkani-Hamed:2002fu,Dvali:2003rk,Nojiri:2003wx,Arkani-Hamed:2003uy,Abdalla:2004sw,Vollick:2003aw,Dvali:2000hr,Deffayet:2001uk}) but also some criticism, partly because they were introduced as purely phenomenological models, but more seriously because it was not clear that they possessed a satisfactory Newtonian limit in the solar system, or that they were free of ghosts (see~\cite{Nunez:2004ts,Chiba:2005nz,Nicolis:2004qq,Luty:2003vm,Koyama:2005tx,Gorbunov:2005zk}). 

In this paper, we investigate the propagating degrees of freedom of the so-called CDDETT model 
\cite{Carroll:2004de}. There already exist detailed studies of the Newtonian limit~\cite{Navarro:2005gh} and the Supernovae contraints~\cite{Mena:2005ta} for this model. Here we derive conditions that they be free of ghosts, and that they have a real propagation speed less than or equal to that of light.

As we review below, a transformation of the action shows that Modified Gravity models are equivalent to a number of scalar fields linearly coupled to higher order curvature invariants.  In the case in which these curvature invariants are fourth order, the relevant one for the Modified Gravity models of Refs.\ \cite{Carroll:2003wy,Carroll:2004de}, we obtain conditions for the propagating degrees of freedom to be well-behaved in their late-time attractor solutions (Friedmann-Robertson Walker spacetimes with accelerating expansion). This extends previous work which established their consistency in de Sitter backgrounds \cite{Nunez:2004ts,Chiba:2005nz,Navarro:2005gh}.

We find that while untroubled by ghosts, the accelerating power-law attractors in general have superluminal tensor and scalar modes, which may place severe theoretical constraints on these models.

\section{The Physical Degrees of Freedom}
Our starting point is the action proposed in~\cite{Carroll:2004de}, which we write in the form
\begin{equation}
S=\int d^4x\,\sqrt{-g}\,[\g\,R-\theta_\mu\,\mu^{4n+2}\,f(Z_i)] \ ,
\label{starta}
\end{equation}
where $\g$ is a constant, $Z_1\equiv R^2$, $Z_2\equiv R_{\a\b}\, R^{\a\b}$ and $Z_3\equiv  R_{\a\b\g\d}\,R^{\a\b\g\d}$. We have introduced $\theta_\mu={\rm sign}(\mu)$ for generality, but note that its presence does not change the late time behaviour of the accelerating attractors, since for an accelerating universe both the $R$ (Einstein-Hilbert) term and the dark matter density become negligible (in other words the exponent of the power law attractor does not depend on $\mu$, see~\cite{Carroll:2004de}). Finally, we take the function $f(Z_i)$ to be of the form
\begin{equation}
f(Z_i)=\frac{1}{(a_i\,Z_i)^n} \ ,
\label{barba}
\end{equation}
where a sum over $i$ is implied.

The action~(\ref{starta}) can be written as that of Einstein gravity coupled to a scalar field, a form more suitable for analysing the propagating degrees of freedom (see the Appendix for a general analysis). Consider
\begin{equation}
S=\int d^4x\,\sqrt{-g}\left[\g\,R-\theta_\mu\,\mu^{4n+2}\left(\frac{n+1}{\phi^n}-\frac n{\phi^{n+1}}\,a_i\,Z_i\right)\right]\ ,
\label{step1}
\end{equation}
where, of course, $\phi\neq0$, otherwise the action is not finite. The variation of this action with respect to 
$\phi$ leads to
\begin{equation}
\phi=a_i\,Z_i\ ,
\end{equation}
and, using this relation, action~(\ref{step1}) and action~(\ref{starta}) yield the same equations of motion. Note that when $a_2=-4\,a_3$ and $a_1=a_3$, this action is equivalent to Einstein-Hilbert gravity coupled to a single scalar through a Gauss-Bonnet (GB) term $R^2_{\rm GB}=R^2-4R_{\mu\nu}R^{\mu\nu} + R_{\mu\nu\rho\si}R^{\mu\nu\rho\si}$.

The coupling of a scalar field with a quadratic expression of the curvature invariants emerges naturally in the context of string theory. In particular, as was shown in~\cite{Gross:1986mw} by Gross and Sloan, in the low-energy effective action the dilaton is coupled to a Gauss-Bonnet term. It is well known that such a term, expanded about a Minkowski vacuum, ensures that the theory is ghost-free (see~\cite{Zwiebach:1985uq}).

It might then seem that taking the $a_i$ to be the GB combination is a sensible choice, because string theory predicts such a coupling to exist and string theory does not have ghosts. However, in models like ours, for which Minkowski spacetime is not a solution, choosing the GB combination of parameters $a_i$ is not a sufficient condition for the non-existence of ghosts.

A ghost is a propagating degree of freedom whose propagator has the wrong sign, and which therefore gives rise to a negative norm state on quantisation. Such states are allowed off-shell in gauge field perturbation theory, but are unacceptable as physical particles.  A theory of gravity with fourth order derivatives 
in the kinetic term inevitably has ghosts \cite{Stelle:1977ry,Barth:1983hb}, but even a 
theory with second order derivatives alone has other potential problems.  Once we break Lorentz invariance, as in a Friedmann-Robertson-Walker (FRW) background, the kinetic terms of a field, even though second order in derivatives, may still have the wrong sign, or may give rise to a propagation speed which is greater than 1, or imaginary. To see this in more detail, consider the action for a scalar field $\phi$,
\bea
S = \int d^4 x \sqrt{-g}\left(  \half T(t) \dot\phi^2 - \half S(t) \nabla\phi^2 \right) \ .
\label{e:problem_action}
\eea
The propagation speed of this scalar 
is $v_\phi = \sqrt{S/T}$.  One may wish to impose one or more of the following conditions
\begin{enumerate}
\item A real propagation speed: $S/T > 0$, otherwise all perturbations have exponentially growing modes.
\item A propagation speed less than light: $S/T < 1$, we will talk about this issue more in detail in section III.
\item No ghosts: $T >0$, to ensure a consistent quantum field theory.
\end{enumerate}
Clearly, unless $S$ and $T$ are positive, and their ratio less than one, we 
will have instabilities, superluminal propagation, or ghosts.  We will see that in studying the 
action for small perturbations of the metric in Modified Gravity theories we will generally encounter 
actions of the form (\ref{e:problem_action}).

If $a_1\neq a_3$, the action~(\ref{starta}) can be written in terms of an Einstein-Hilbert term plus a particular extra piece involving two new scalar fields. Furthermore, because of the special properties of the Gauss-Bonnet term, the equations of motion are no longer 4th order, but remain 2nd order in the fields. Taking the action~(\ref{step1}) and introducing a new scalar field $\sigma$, we have
\bea
S=\int d^4 x\sqrt{-g}\,\left\{[\g+2b\,\sigma\,f(\phi)]\,R-U(\phi)-b\sigma^2f(\phi)+f(\phi)\GB\right\}
\label{HN1}\ ,
\eea
where
\bea
U(\phi)&=&\theta_\mu\,\mu^{4n+2}\,\frac{n+1}{\phi^n}\\
f(\phi)&=&a_3\,\theta_\mu\,\mu^{4n+2}\,\frac n{\phi^{n+1}}\ ,
\eea
with $b\equiv a_1/a_3-1$, and $\GB$ the Gauss-Bonnet invariant. 

Making a field redefinition $\sigma\equiv\chi/(2bf)$, the equation of motion for $\chi$ is $\chi=2bfR$, and the gravitational equations then become
\bea
&(\g+\chi)\,R_{\a\b} - \N_\a\N_\b\chi+g_{\a\b}\,\Box\chi-2\,R\,\N_\a\N_\b f 
+ 2\,g_{\a\b}\,R\,\Box f +8R_{(\a\nu}\,\N_{\b)}\N^\nu f & \nonumber \\
&-4\,R_{\a\b}\,\Box f-4\,g_{\a\b}\,R^{\r\s}\N_\r \N_\s f
- 4\,R_{(\a}{}^{\s\t}{}_{\b)}\,\N_\s\N_\t f-\tfrac12\,g_{\a\b}\left[(\g+\chi)\,R-\frac{\chi^2}{4\,b\,f}-U\right]=0 \ .&
\eea
Independent of the background, after application of the Bianchi indentities these equations are second-order in derivatives of the fields, thanks to the Gauss-Bonnet combination. 

It is known that adding terms quadratic in the curvature invariants to the Einstein-Hilbert action with a cosmological constant yields an extra scalar mode and a spin-2 mode, which is generically a ghost because of its fourth-order field equation (see~\cite{Stelle:1977ry,Hindawi:1995an,Boulanger:2000rq,Woodard:2006nt}). Thus, provided we are expanding around a constant $\phi$ background, such as de Sitter space, we can directly infer that Modifed Gravity models are generically afflicted by spin-2 ghosts. However, in our case, it is clear that the higher derivative terms cancel out identically because $a_2=-4\,a_3$, as was already found in~\cite{Nunez:2004ts,Chiba:2005nz,Navarro:2005da}.  The remaining extra scalar degree of freedom, $\chi$, is due to the presence of the extra $R^2$ term in the lagrangian, which vanishes if $a_1=a_3$.

Crucially though, the vanishing of the fourth order term is a necessary but not sufficient condition for the 
absence of ghosts in the spin-2 sector. As we described above, one must also separately check the signs of the second order derivatives with respect to both time and space (a check was not performed in \cite{Navarro:2005da}). 

In this paper we derive and study the kinetic terms for both the spin 0 and spin 2 fields in time-dependent backgrounds, to which derivatives of $f$ contribute. We find that both fields may be afflicted by instabilities or ghosts, contrary to the claim of the absence of ghosts in FRW spacetimes made in \cite{Navarro:2005da}. We show that the special case of an empty accelerating universe  -- the late-time attractor of an FRW cosmology in the Modified Gravity model under consideration -- the propagating states are generically (but not universally) superluminal over the $(b,n)$ parameter space.

\section{Propagation in FRW spacetimes}
If the second order derivatives of the spin-2 and spin-0 fields do not have the correct signs in FRW spacetimes, then the theory may be inconsistent. The existence of a ghost mode would lead, for example, to the over-production of all particles coupled to it. One may think of a theory with ghosts as an effective theory, with no ghosts above some cutoff, thereby restoring consistency (see~\cite{Carroll:2003st,Cline:2003gs}). However, this cut-off must be less than about 3 MeV \cite{Cline:2003gs}. 

A further condition that one may wish to impose is that the propagation speeds be less than or equal to unity. One worry is that the existence of superluminal modes on the relevant cosmological backgrounds may lead to a catastrophic signature of causality violation (see e.g.~\cite{Adams:2006sv}).  Other authors~\cite{Babichev:2006vx,Rendall:2005fv,Vikman:2006hk}) have discussed the problem of superluminal propagation in non-lorentzian backgrounds and, in particular, have suggested that the presence of superluminal modes would introduce a second horizon, the so called sound horizon, different from the light causal horizon, which may lead to ambiguities and inconsistencies in black-hole thermodynamics~\cite{Dubovsky:2006vk}. Furthermore, it has been pointed out (\cite{Adams:2006sv,Rendall:2005fv,Armendariz-Picon:2005nz}) that for some set of initial conditions, superluminal modes may yield ill-posed Cauchy problems. In particular, in our case, nothing prevents $c_0^2$ and~$c_2^2$ - the respective speeds of the scalar and tensor modes - becoming infinite. In general this would lead to causally connected spatial sections and eventually to an ill-posed Cauchy problem. 

On the other hand, in the context of non-commutative geometry, other authors have studied superluminal propagation and shown that there is no causality violation if there is a preferred reference frame~\cite{Hashimoto:2000ys}.

Given these different possibilities, in this paper we will present the constraints from superluminal propagation in a clearly distinct way from those arising from ghosts, so as to allow readers to impose fewer or more constraints, depending on the particular theory they are working with.

\subsection{Pure Modified Gauss-Bonnet Gravity}

In this section we begin with the special case 
\begin{equation}
S = \int d^4x\sqrt{-g}\,
\left(\gamma R + f(\phi) \GB - U(\phi) \right),
\end{equation}
which we refer to as the Modified GB action. We vary with respect to $\phi$ and $g_{\al\be}$, 
write $\delta g_{\al\be} = h_{\al\be}$, so that $\delta g^{\al\be} = -h^{\al\be}$, and use the many 
useful identities contained in Ref.\ \cite{Barth:1983hb} to derive 
\begin{eqnarray}
\de_\phi S  & = & 
\int d^4x\sqrt{-g}\, 
\left( 
\GB f'(\phi)- U'(\phi)
\right)\de\phi \\
\de_g S & = & 
\int d^4x\sqrt{-g}\, 
\left\{ 
\half R \left[ \left( \ga - 4 \, \Box f \right)g^{\al\be}  + 4\na^\be\na^\al f  \right]\right. \nonumber\\
&& - R^{\mu\nu} \left[ \left( \ga - 4 \, \Box f \right)\de^\al_\mu\de^\be_\nu - 4 (\na_\nu\na_\mu f) g^{\al\be} 
+ 8( \na^\rh\na_\mu f) \de^{(\al}_\rh\de^{\be)}_\nu \right] \nonumber\\
&& \left. + {{R^\al}_{\nu\rho}}^\be (4 \na^\rh\na^\nu f) - \half U(\phi)g^{\al\be}
\right\} h_{\al\be} \ .
\label{e:deltag}
\end{eqnarray}
In order to establish whether the theory is stable and ghost free, we must examine the second variation of the action. This can be organized as before, 
\begin{equation}
\de^2S = \de_\phi^2 S + 2\de_g\de_\phi S + \de_g^2S \ .
\end{equation}
The easiest term is
\begin{equation}
\de^2_\phi S = \int d^4x\sqrt{-g}\, 
\left( 
\GB f''(\phi)- U''(\phi)
\right)\de^2\phi \ ,
\end{equation}
and we can simplify the mixed term using the field equation for $\phi$, to obtain
$\de_g\de_\phi S = \int d^4x\sqrt{-g}\, 
(\de_g\GB) \de f$, where $\de f=f'(\phi)\de\phi$.  After integration by parts we then have
\begin{eqnarray}
\de_g\de_\phi S &=&  \int d^4x\sqrt{-g}\, 
h_{\al\be}\left[
(4R^{\al\be} - 2R g^{\al\be})g^{\rho\si}  \right.\nonumber\\
&& \left.+(2Rg^{\be\rh}g^{\al\si} + 4 R^{\rh\si} g^{\al\be} - 
8 R^{\si(\be} g^{\al)\rh}) + 4 R^{\al\si\rh\be} 
\right]  \na_\rh\na_\si\de f \ .
\end{eqnarray}
We simplify the final term using the field equation for $g_{\al\be}$, and organise according to the 
number of derivatives of $h_{\al\be}$.  In order to check that the original action is ghost free, we need to 
establish that the fourth order terms vanish, and that terms involving two derivatives of the metric have the appropriate sign.

It is already straightforward to see from Eq.\ (\ref{e:deltag}) that there can be no fourth order 
derivatives, as the terms containing derivatives of the Riemann and Ricci tensors, and the Ricci 
scalar, have already cancelled.  The remaining second order terms are
\bea
\de_g^2 S_{(2)} & = &  \int d^4x\sqrt{-g}\, 
h_{\al\be}\left\{
-\quarter \left[ (\ga - 4\Box f)g^{\al\be} + 4\na^\be\na^\al f\right]
\left( \Box h + \na^\de\na^\ga h_{\ga\de}\right)\right.\nonumber\\
&+&  \half\left[(\ga - 4\Box f)\de^{\al}_\mu\de^{\be}_\nu  - (4 \na_\nu\na_\mu f) g^{\al\be} 
+ (8 \na^\rh\na_\mu f) \de^{(\al}_\rh \de^{\be)}_\nu \right]\notag\\
&&\quad{} \times(\Box h^{\mu\nu} + \na^\nu\na^\mu h - \na^\mu A^\nu - \na^\nu A^\mu)
\nonumber\\
&+& \left. 
\half\left[4\na_\rh\na_\nu f\right]\left( \na^\rho\na^\nu h^{\al\be} + \na^\be\na^\al h^{\nu\rh}
       - \na^\be\na^\rh h^{\al\nu} - \na^\be\na^\nu h^{\al\rh}  \right)
\right\},
\eea
where $A^\mu = \na_\nu h^{\mu\nu}$.

In an FRW space-time, the background fields break Lorentz invariance, with 
\bea
g_{\mu\nu} &=& a^2(\tau)\eta_{\mu\nu}, \\
\na_\nu\na_\mu f & = & \de^0_\mu\de^0_\nu B(\tau) + g_{\mu\nu} g^{00}C(\tau),
\eea
where $B = f'' - 2\He f'$, $C = \He f'$ (and where we have assumed spatial flatness). Here a
prime denotes a derivative with respect to conformal time $\tau$, and $\He \equiv a'/a$. 
In this background 
it is convenient to decompose the metric perturbation into scalar, vector and tensor modes 
in the usual way. To check the sign of the kinetic term for the spin-2 particle, we first identify it in 
the expansion of the metric
\bea
h_{ij} = g^{(3)}_{ij}\varphi + \ga_{|ij} + C_{(i|j)} + H_{ij},
\eea
(where the symbol $|$ denotes covariant differentiation with respect to the spatial metric $g^{(3)}_{ij}$)
as the transverse, traceless (${H^i}_{j|i}=0$, ${H^i}_{i} = 0$) tensor mode $H_{ij}$.  It is straightforward to show that the tensor part of the second 
variation which is second order in derivatives of $h_{\mu\nu}$ is
\bea
\de_g^2 S^T_{(2)} & = &  \int d^4x\sqrt{-g}\,
\left[
\half H_{ij} \left(\ga - 4 g^{00}(B+C) \right) \Box H^{ij} + \half H_{ij} (4B) \na^0\na^0H^{ij}
\right].
\eea
Rearranging and using $\Box = \na_0\na^0 + \triangle$ to display the time and space derivatives separately, we find
\bea
\de_g^2 S^T_{(2)} & = &  \int d^4x\sqrt{-g}\,
\left[
\half H_{ij} \left(\ga - 4 g^{00}(B+C) \right) \triangle H^{ij} + \half H_{ij} (\ga -4g^{00}C) \na_0\na^0H^{ij}
\right].
\eea
We therefore have two conditions for a stable theory free of ghosts:
\bea
\ga + \frac{4}{a^2}(f'' - \He f')>0, \qquad \ga+ \frac{4}{a^2}\He f' > 0 \ .
\label{condeta}
\eea

The ratio of the coefficients of the second derivatives is the propagation velocity squared of the spin-2 mode.
Therefore, in terms of physical time, $t = \int a(\tau)d\tau$ and Planck units $\ga = 1/2$, the condition that a background have a real and non-superluminal spin-2 propagation speed is
\bea
0\le c_2^2 = \frac{1 + 8 \ddot f}{1+ 8H\dot f } \le 1 \ ,
\label{Spin2Speed}
\label{cond1}
\eea
where a dot indicates a derivative with respect to $t$, and $H\equiv \dot a/a$. Further, the condition that the spin-2 mode not be a ghost is 
\bea
0 < 1 + 8 H\dot f,\label{cond2} \ . 
\eea

The same strictures apply to the scalar (spin-0) mode, whose kinetic term is much more difficult 
to evaluate.  Fortunately, our Lagrangian is a special case of a class of theories studied in 
Ref. \cite{Hwang:1999gf}.  There it was shown that the gauge-invariant combination $\Phi = \varphi - H\de\phi/\dot\phi$ (one of the Bardeen scalars) satisfies
\bea
\frac{1}{a^3}\frac{\pa}{\pa t}\left( a^3 Q \dot \Phi \right) - P \frac{\triangle}{a^2}\Phi =0,
\eea
where
\bea
Q & = & \frac{96H^2\dot f^2(1+8H\dot f)}{(1+12H\dot f)^2},\\
P & = & \frac{32H^2\dot f^2}{(1+12H\dot f)^2}
\left[-8\ddot f+8H\dot f+(1+8H\dot f)\left(3+4\frac{\dot H}{H^2}\right)  \right] \ .
\eea

If the spin-2 propagator is well behaved, the sign of the scalar time derivatives is also correct thanks to Eq.\ (\ref{cond2}), and there are no spin-0 ghosts. Finally, the condition that the scalar propagation speed be real and non-superluminal reads
\bea
0\le c_0^2 = \left(1+\frac43\frac{\dot H}{H^2}\right)-\frac83\frac{(\ddot f-H\dot f)}{(1+8H\dot f)} \le 1 \ .
\label{Spin0Speed}
\label{cond3}
\eea

Having established these results, let us now focus on the particular class of models in hand.
We have
\bea
f=\frac A{24}\,\frac1{\phi^{n+1}}\ ,
\eea
where $A=24\,n\,\theta_\mu\,\mu^{4n+2}$. Without loss of generality, we fix to unity the coefficient of the square of the Riemann tensor in the lagrangian, i.e.~$a_3=1$, and note that the $\phi$ equation of motion, in a flat FRW background, then yields
\bea
\phi=\GB=24\,H^2\,\frac{\ddot a}a\ .
\eea
Since we are interested in the behaviour of these actions in the universe at late times, we study the attractor solutions for the CDDETT model when the matter is diluted away because of the expansion. In general~\cite{Carroll:2004de} the attractor solutions can be written as $a(t)\propto t^p$, with $p>1$. For the Gauss-Bonnet combination, it was found that the relevant accelerating power-law attractor is given by $p=4n+1$. 
We see that
\bea
\frac\phi{24}=(p-1)\,\frac{p^3}{t^4}\ ,
\eea
with $\phi>0$ for an accelerating universe, and therefore
\bea
\dot f &=&4\,(n+1)\,(p-1)\,\frac A{\phi^{n+2}}\,\frac{p^3}{t^5}\ ,\\
\ddot f&=&96\,(n+1)\,(4n+3)\,(p-1)^2\,\frac A{\phi^{n+3}}\,\frac{p^6}{t^{10}}=\frac{4n+3}t\,\dot f\ .
\eea
Thus, it is clear that, if $A>0$ or $\theta_\mu=1$, (\ref{condeta}) are satisfied for an accelerating universe ($p>1$). However, the spin-2 propagation speed is
\bea
c_2^2 = \frac{1 + 8(4n+3)\dot f/t}{1 + 8p\dot f/t},
\eea
which for the pure GB theory approaches 1 from above, and so the graviton propagates faster than light.

For the scalar, condition (\ref{cond3}) amounts to 
\bea
0\le c_0^ 2 = \left( 1 - \frac{4}{3p}\right) + \frac83 \frac{(p - 4n-3)\dot f/t}{1 + 8p\dot f/t} \le 1 \ .
\eea
A sufficient condition to satisfy these relations is 
\bea
\max(4,n+7/4) < p < 4n+3 \ ,
\eea
which is clearly satisfied for the pure GB theory for any $n\geq1$. 

To summarise: there are no ghosts or instabilities for the Modified GB attractor solutions for any $n\geq1$\footnote{Note that the vector modes do not propagate \cite{Hwang:1999gf}}.  However, the graviton propagates superluminally, which may render the pure GB theory inconsistent. 

Finally, the pure GB combination is not phenomenologically viable. In this case, $\phi\propto H^2\ddot a$ and $\phi$ vanishes as $\ddot a$ approaches to zero. This means that, for this combination, in the CDDETT model $\ddot a=0$ is a singularity of the equations of motion, and it not possible to change the sign of $\ddot a$. The Universe can never change from deceleration to acceleration. (This is reminiscent of the problem encountered in some other modified gravity theories~\cite{Easson:2005ax}.) 

\subsection{Modified Gravity with matter}

In this section we study the propagation of the spin-2 and spin-0 modes
when matter is present, which changes the background cosmology and hence the coefficients of the second derivatives in the action. 

As we discovered in the previous subsection, a realistic model should have $a_1\neq a_3$ so that the modification is not pure Gauss-Bonnet. The conditions (\ref{cond1},\ref{cond2},\ref{Spin2Speed},\ref{cond3},\ref{Spin0Speed}) can be generalized in a straightforward but lengthy way (see~\cite{Hwang:2005hb}), and depend on the parameter $b=a_1/a_3-1$ (recall we are assuming that $a_2 = -4a_3$ as required to cancel fourth-order derivatives). The correct signs in the spin-2 propagator are assured if
\bea
1 + 4bfR  +8 \ddot f &>&0,\label{cond_gen1} \label{e:RealTensorSpeedMatter}\\
1+ 4bfR+8H\dot f &>& 0,\label{cond_gen2} \label{e:noGhostMatter}
\eea
with the propagation speed condition reading
\bea
c_2^2 = 
\frac{1 + 4bfR  +8 \ddot f }{1+ 4bfR+8H\dot f} \le 1.
\label{e:SublumGraviton}
\eea
The coupling of $f(\phi)$ to the Ricci scalar outside the GB combination gives an extra propagating scalar degree of freedom 
$\Psi$  (see~\cite{Hwang:2005hb}), the other Bardeen scalar.
The two scalar modes have the same speed of propagation $c_0^2$ and, once again, if the graviton propagator is well behaved, there is only the following extra condition to be satisfied
\bea
0\leq c_0^2=1+\frac{32}{3Q_1}\,\dot f\dot H-\frac{8}{3Q_2}(\ddot f-\dot fH),
\label{cond_gen3}
\eea
where
\bea
Q_1=4b(\dot fR+f\dot R)+8\,\dot fH^2\qquad{\rm and}\qquad
Q_2=1+4bfR+8H\dot f\ .
\eea
These conditions put additional bounds on the parameter space spanned by the $\mu,a_i$ which define the CDDETT Modified Gravity theories, if we require that they hold at all times during the evolution of the universe. In particular, we may require that they hold when the Universe expands with a power-law, $a\propto t^p$, for which 
\bea
\phi&=&\frac{\Gamma_1}{t^4}\\
R&=&\frac{\Gamma_2}{t^2}\\
f&=&\Gamma_3\,t^{4n+4}\ .
\eea
where
\bea
\Gamma_1&=&12p^2[3b(2p-1)^2+2p(p-1)]\\
\Gamma_2&=&6p(2p-1)\\
\Gamma_3&=&\frac{n\,\theta_\mu\,\mu^{4n+2}}{a_3^n\,\Gamma_1^{n+1}}\ .
\eea
Therefore we have that
\bea
c_2^2&=&\frac{1+4\,\Gamma_3\,t^{4n+2}\,[b\Gamma_2+8(n+1)(4n+3)]}%
{1+4\,\Gamma_3\,t^{4n+2}\,[b\Gamma_2+8p(n+1)]}, 
\label{latet2}\\
c_0^2 &=& 1 - 
  \frac{32}{3}\frac{p(n+1)}{b\Ga_2(4n+2)+8p^2(n+1)} -
  \frac{8}{3}\frac{(n+1)(4n+3-p)}{[b\Ga_2 + 8p(n+1)+1/4\Ga_3t^{4n+2}]}
\label{latet0}
\eea
This relation holds for any power-law behaviour, unless either $\Gamma_1$ or $\Gamma_2$ becomes zero. We expect power-law expansion at early time, where the Universe should behave as an ordinary Friedmann model, and at late time where it approaches an accelerating attractor solution. 

Let us consider a matter-dominated Universe,  $p=2/3$, assuming a positive $\Gamma_3$ to ensure the no-ghost conditions are satisfied. In this case it is clear that the tensor modes are superluminal, and their speed tends to one as $t\to 0$.

On the other hand, for a late-time attractor solution,  the accelerating power law is given by \cite{Carroll:2004de}
\bea
p=\frac{2(n+1)(4n+1)-3\alpha+\sqrt{9n^2\alpha^2-4(n+1)(4n+1)(5n+1)\alpha+4(4n+1)^2(n+1)^2}}{4(n+1)} \ ,
\eea
where $\alpha=(12a_1+4a_2+4a_3)/(12a_1+3a_2+2a_3)=6b/(6b+1)$. The exponent $p$ is real, for large $n$, if $\al\lesssim\tfrac89\, n$ or $\al\gtrsim 8\,n$. Furthermore at late times, Eq.~(\ref{latet2}) takes the approximate form
\bea
c_2^2\approx\frac{b\Gamma_2+8(n+1)(4n+3)}{b\Gamma_2+8p(n+1)}\ ,
\eea
which means that for non-superluminal behaviour we require
\bea
p>4n+3\ .
\eea
It should be noted that the value $\alpha=1$ cannot be considered, because this would imply that $a_2\neq-4a_3$. For large $n$, one requires
\bea
\al \gtrsim \frac{80n}{9}, \quad \mathrm{or} \quad \al \lesssim  -\frac85
\eea
to avoid superluminal tensor modes.

At late times, Eq.~(\ref{latet0}) takes the approximate form
\bea
c_0^2 &=& 1 -
  \frac{32}{3}\frac{p(n+1)}{b\Ga_2(4n+2)+8p^2(n+1)} -
  \frac{8}{3}\frac{(n+1)(4n+3-p)}{b\Ga_2 + 8p(n+1)},
\label{scalla}
\eea
For large $n$, Eq.\ (\ref{scalla}), implies that the scalar modes are not superluminal if
\bea
-\frac{24}5\lesssim \al\lesssim1\ .
\eea
For large $n$, in the region $-24/5\lesssim\alpha\lesssim-8/5$ both tensor and scalar modes are not superluminal. In the case of $n=1$ one can see that the allowed region is $-3.793\lesssim\al\leq-\frac{112}{57}\approx-1.965$, where the lower bound is given by $c_2^2=0$ and the upper one by $c_2^2=1$. This additional constraint would rule out the region $0.9<\alpha<1$ (where $p$ is real) identifed by Mena at al.\ as producing cosmologies consistent with Supernovae data. 

\begin{figure}[ht]
\begin{center}
{\psfrag{n}[][][1][-90]{$n$}
\psfrag{a}{$\al$}
\includegraphics[width=8truecm]{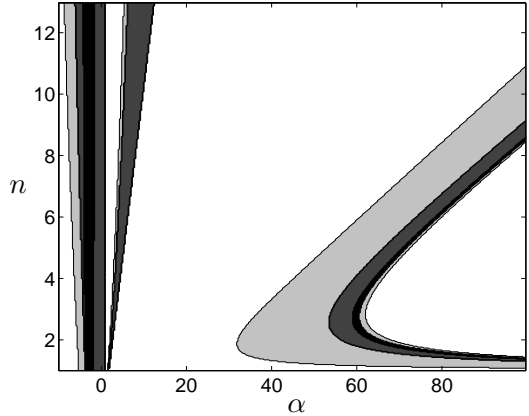}}
\label{f:guarda}
\caption{Contour plot in the $(\al,n)$ plane for the constraints. The light grey area corresponds to the region in which only the no-ghost constraint (Eq.\ \ref{e:noGhostMatter}) holds. The darker area represents the points at which both the no-ghost and the positive-squared-velocity conditions 
(Eqs.\ \ref{e:RealTensorSpeedMatter}, \ref{e:noGhostMatter}, \ref{cond_gen3})
hold at the same time. Finally the darkest region is the region of the plane at which all the constraints (no-ghost, $0<c^2<1$, Eqs.\ \ref{e:RealTensorSpeedMatter}, \ref{e:noGhostMatter}, \ref{cond_gen3}, \ref{e:SublumGraviton})
hold for both scalar and tensor modes.}
\end{center}
\end{figure}

A plot of the allowed values of $\al$ and $n$, after imposing the no-ghost constraint, or the no-superluminal constraint or both, is shown in Fig.\ \ref{f:guarda}. 

\section{Conclusions}
The search for a satisfactory model for the acceleration of the universe has been pursued in many different ways. Recently, models attempting to explain such behavior by changing the gravity sector have been proposed~\cite{Carroll:2003wy,Carroll:2004de,Capozziello:2003tk,Deffayet:2001pu,Freese:2002sq,Arkani-Hamed:2002fu,Dvali:2003rk,Nojiri:2003wx,Arkani-Hamed:2003uy,Abdalla:2004sw,Vollick:2003aw,Dvali:2000hr,Deffayet:2001uk}. In particular, the CDDETT model~\cite{Carroll:2004de}, has the attractive feature of the existence of accelerating late-time power-law attractors, while satisfying solar system constraints \cite{Navarro:2005da}.

In this paper we have investigated the consistency of the propagating modes (tensor and scalar) for the action
\begin{equation}
S=\int d^4x\,\sqrt{-g}\,\left[\g\,R+\theta_\mu\,
\frac{\mu^{4n+2}}{(a_1\,R^2+a_2\,R_{\a\b}\,R^{\a\b}+a_3\,R_{\a\b\g\d}\,R^{\a\b\g\d})^n}\right]\ .
\end{equation}
In order for this action to be ghost-free, it is necessary but not sufficient, to set $a_2=-4\,a_3$ \cite{Chiba:2005nz,Navarro:2005da} so that there are no fourth derivatives in the linearised field equations. What remained was the possibility that the second derivatives might have the wrong signs, and also might allow superluminal propagation at some time in a particular cosmological background. For example, for the case $a_1=a_3$, for which the modification is a function of the Gauss-Bonnet term,  we found that the accelerating power-law attractor solutions give propagators with the correct signs, but with a spin-2 mode propagating faster than light.

We have also examined the general second order CDDETT Modified Gravity theory in a FRW background with matter, which is parametrized by the energy scale $\mu$ and by $b = a_1/a_3 - 1$ - the deviation of the Ricci scalar-squared term from that appearing in the Gauss-Bonnet combination - or equivalently $\al = 6b/(6b+1)$. We found that the theories are ghost-free, but contain superluminally propagating scalar or tensor modes over a wide range of parameter space. In conclusion, we note that there are likely to be further constraints from compatibility with CMB data as we have changed gravity on large scales quite significantly.  To investigate this point is beyond the remit of the current paper.

\begin{acknowledgments}
We want to thank Gianluca Calcagni, Sean Carroll, Gia Dvali, Renata Kallosh, Kei-ichi Maeda, Shinji Mukhoyama, Burt Ovrut, Paul Saffin, Karel Van Acoleyen and Richard Woodard for helpful comments and discussions. ADF is supported by PPARC. MT is supported in part by the NSF under grant PHY-0354990 and by Research Corporation.

\end{acknowledgments}

\appendix

\section{Scalar fields and Modified Gravity}

\label{a:scalars}

In this Appendix we demonstrate how actions for non-standard models of gravity can be rewritten in the form of Einstein 
gravity with a non-minimal coupling to one or more scalar fields. This is not a new result (see most recently e.g. \cite{Chiba:2005nz}), but the Modified Gravity action (\ref{barba}) is a special case which needs separate consideration.

Consider the action
\begin{equation}
S=\int d^4x\,\sqrt{-g}\,F(Z_i)\ ,
\label{GenAct}
\end{equation}
where $Z_i$ are monomials in the curvature invariants, with $Z_1$, $Z_2$ and $Z_3$ defined earlier, but where we allow the possibility of higher order terms.  If the function $F(Z_i)$ is at least twice differentiable (except possibly at isolated points), then this is easily seen to be equivalent to the action 
\begin{equation}
S_{\phi_i}=\int d^4x\,\sqrt{-g}\,[F(\phi) + (Z_i - \phi_i)F_i(\phi)] \ ,
\label{ScalarGenAct}
\end{equation}
where $\phi_i$ are a set of auxiliary scalar fields, one for each of the terms $Z_i$, 
and $F_i \equiv \partial_iF$. The first variation is 
\begin{equation}
\delta S_{\phi_i}  = \int d^4x\,\sqrt{-g}\,[(Z_i - \phi_i)F_{ij}(\phi) \delta \phi_j + \delta Z_i F_i(\phi) + \half \lag_{\phi_i}\delta g ],
\label{first Var}
\end{equation}
where $\lag_{\phi_i} = F(\phi) + (Z_i - \phi_i)F_i(\phi)$.  We immediately see that, provided the matrix of second derivatives $F_{ij} = \partial_i\partial_jF(\phi)$ is non-singular, $\phi_i = Z_i$, and we return to the original action (\ref{GenAct}).   This was one of the results in \cite{Chiba:2005nz}.

However, the possibility of a singular matrix was not considered. In models of  the form 
(\ref{barba}), there exist degeneracies in the parameters of the form
\begin{equation}
F(Z_i - A_\alpha v^\alpha_i) = F(Z_i) \ ,
\label{degen}
\end{equation}
where $A_\alpha$ are arbitrary constants, and $v^\alpha_i$ are orthonormal vectors in the space of curvature invariants.  In this case, $v_i^\alpha F_i =0$, $F_{ij}v^\alpha_j=0$, and higher 
derivatives also vanish.  

The solution of $\delta S/\delta\phi_i = 0$ is now 
\begin{equation}
(Z_i - \phi_i) = A_\alpha v_i^\alpha \ ,
\end{equation}
with $A_\alpha$ again arbitrary constants. By substitution we find  
\begin{equation}
S_A=\int d^4x\,\sqrt{-g}\,[F(Z_i -  A_\alpha v_i^\alpha) 
+ A_\alpha v_i^\alpha\partial_iF(\phi)] = \int d^4x\,\sqrt{-g}\,F(Z_i) \ ,
\end{equation}
which is again our original action.  

As our Modified Gravity example suggests, we can reduce the number of scalar fields by taking linear combinations of the $Z_i$ normal to the subspace spanned by the $v^\alpha_i$.  If this subspace is spanned by $w^A_i$, then we can define a new set of variables $Z^{\prime}_A  = w^A_iZ_i$, $Z^{\prime}_{\alpha} = v^\alpha_iZ_i$. 
The function $F$ is now independent of $Z^{\prime}_{\alpha}$, so we can write
\begin{equation}
S[g_{\mu\nu},\phi_A]=\int d^4x\,\sqrt{-g}\,[F(\phi'_A) + (Z^{\prime}_{A} - \phi'_A)F_A(\phi')] \ .
\end{equation}
In the CDDETT model, the Lagrangian density depends on $Z_1$, $Z_2$, and $Z_3$ only through the 
combinations $Z_1$, $ - 4 Z_2 + Z_3$. There is one degeneracy, and so there are only two scalar fields required to put the action into the linearised form of Eq.\ (\ref{ScalarGenAct}).  When $b=0$ in (\ref{HN1}), the field associated with $Z_1=R^2$ may be trivially solved for to give the Einstein-Hilbert term.

\end{document}